\documentstyle[12pt]{article}

\newcommand{\com}[2]{\left[ #1,#2 \right]}               
\newcommand{\acom}[2]{ \{ #1,#2 \} }                     
\newcommand{\cyclicsum}[3]{\oint_{#1,#2,#3}}             
\newcommand{\resetcounter}{\setcounter{equation}{0}}     

\newcommand{\D}  {{\cal D}}               

\newcommand{\A}  {\alpha}
\newcommand{\B}  {\beta}
\newcommand{\E}  {\varepsilon}

\newcommand{\Sm} {\sigma}

\newcommand{\C} {\gamma}

\begin{document}

\thispagestyle{empty}
\begin{titlepage}
\begin{flushright}
HUB--EP--95/34 \\
hep-th/9512165 \\
\end{flushright}
\vspace{0.3cm}
\begin{center}
\Large \bf Extended supersymmetry with
           \\ gauged central charge
\end{center}
\vspace{0.5cm}
\begin{center}
Ingo \ Gaida$^{\hbox{\footnotesize{1}}}$,  \\
{\sl Institut f\"ur Physik, Humboldt--Universit\"at,\\
 Invalidenstrasse 110, D--10115 Berlin, Germany}
\end{center}
\vspace{0.6cm}

\begin{abstract}
\noindent
Global $N=2$ supersymmetry in four dimensions with a gauged central
charge is formulated in superspace.
To find an irreducible representation of supersymmetry for
the gauge connections a set of constraints is given.
Then the Bianchi identities are solved subject to this
set of constraints. It is shown that the gauge
connection of the central charge is a $N=2$ vector multiplet.
Moreover the Bogomol'nyi bound of the massive particle states
is studied.
\end{abstract}

\vspace{0.3cm}
\footnotetext[1]{e-mail: gaida@qft2.physik.hu-berlin.de,
                 supported by Cusanuswerk}
\vfill
\end{titlepage}


\setcounter{page}{1}

%
%

\resetcounter

\setcounter{section}{1}

In this paper, global $N=2$ supersymmetric theories in four dimensions
are discussed. Although these theories are not directly related to
physics at the electroweak scale, they have been extensively
studied in the last years.
It started with an analysis of the vacuum structure of
$N=2$ supersymmetric gauge theories with gauge group
$G = SU(2)$ by Seiberg and Witten, who showed that these theories
exhibit physical phenomena like asymptotic freedom, chiral symmetry
breaking and confinement of the electric charge
[\ref{Seiberg_Witten}]. Moreover a version
of Olive-Montonen electric-magnetic duality appears
[\ref{Montonen_Olive}].
Their results have been extended to other gauge groups
[\ref{Theisen_Klemm},\ref{Argyres_Faraggi}] and to the case of
effective string theories in four dimensions
[\ref{Luest},\ref{Antoniadis}].
\\
In the context of the electric-magnetic duality the central
charge plays an important role, because the electric and the magnetic
charges can appear as a complex central charge in the algebra.
Technically these central charges enter a supersymmetric theory
because certain surface terms, usually discarded in deriving the algebra,
are non-vanishing [\ref{Witten_Olive}].
\\
The generator of the central charge
is defined to commute with all the other
generators of the theory [\ref{Haag}]. From the point of view of the
$N=2$  supersymmetry algebra the appearance of a central charge
leads to new torsion terms. As a consequence
the Bianchi identities differ from the ones without
central charge. The solution of the latter is well known
[\ref{Grimm_Wess_Sohnius}].
\\
In this paper the solution of the Bianchi identities with gauged central
charge will be given in the context
of global $N=2$ supersymmetry.
In $N=2$ supergravity the central charge must be gauged anyway
[\ref{gauged_central_charge}].
To solve the Bianchi identities
one has to find an appropriate set of constraints. These
constraints yield an irreducible representation of supersymmetry
for the $U(1)$ gauge connections.
\\
It is well known that the central charge can determine
the quantum mechanical mass spectrum by the use of the
Bogomol'nyi bound [\ref{Witten_Olive}].
It is shown that
the gauging of the central charge can change this mass bound, if
the lowest component of the gauge connection of the central charge
(Higgs field) has a non-vanishing vacuum expectation value.

\vspace{1cm}

The paper is organized as follows:
First of all the $N=2$ supersymmetry algebra
with gauged central charge is introduced.
For simplicity only the case where the
central charge is gauged by one $U(1)$ gauge group is considered.
Then the set of constraints on the field strengths
is given and the solution of the Bianchi identities subject to
these constraints is shown.
In particular the gauge connection of the central charge is
discussed in detail.
Moreover the possible Higgs-effect, which is a consequence of
the gauging of the central charge, is studied.
The whole discussion is given in the framework of rigid
$N=2$ superspace.

\vspace{1cm}


After the formulation of $N=1$ supersymmetric theories
an extended version with two supersymmetry generators has been formulated
[\ref{Ferrara_Zumino},\ref{Firth_Jenkins},\ref{Fayet}].
The corresponding algebra is\footnote{
Here the following conventions for complex conjugation
and for the $SU(2)_{R}$ metric $g_{ij}$ is used:
 $\D_{\A}^{ \ i \ +} =  \bar \D_{\dot\A  i},
   g^{12} = 1,
   g^{ij} = - \ g^{ji} \ = \ -g_{ij},
   g^{ij} \ g_{jk} = \delta^{i}_{ \ k}$. Moreover the conventions of
   [\ref{Wess_Bagger}] are used.
}
\begin{eqnarray}
\label{n=2ungauged_susy_algebra2}
 \acom{ \D_{\A}^{ \ i} }{ \bar D_{\dot\A j} }
           &=& -2 i \ \delta^{i}_{ \ j} \ \Sm_{ \ \A \dot\A}^{m} \ \partial_{m}
\nonumber\\
 \acom{ \D_{\A}^{ \ i} }{ \D_{\B}^{ \ j} }
                            &=& -2i \ g^{ij} \ \E_{\A\B} \ \partial_{z}
\nonumber\\
\acom{ \bar \D_{\dot\A i} }{ \bar \D_{\dot\B j} }
      &=& -2i \ g_{ij}  \ \E_{\dot\A\dot\B} \ \partial_{\bar z}
\end{eqnarray}

All the other graded commutators vanish.
The occurence of the central charge leads to the new torsion
$T_{\A\B}^{ \ \ z ij} = 2 \ g^{ij} \ \E_{\A\B}$ in 
(\ref{n=2ungauged_susy_algebra2}).
Note that the central charge index $z$ is used as an internal index.
Gauging this algebra with an abelian gauge group yields

\begin{eqnarray}
\label{n=2susy_algebra2}
 \acom{ \D_{\A}^{ \ i} }{ \bar D_{\dot\A j} }
           &=& -2 i \ \delta^{i}_{ \ j} \ \Sm_{ \ \A \dot\A}^{m} \ \D_{m}
               \ + \ i \  F_{\A\dot\B \ \ \ j}^{ \ \ \ i}
\nonumber\\
 \acom{ \D_{\A}^{ \ i} }{ \D_{\B}^{ \ j} }
                            &=& -2i \ g^{ij} \ \E_{\A\B} \ \D_{z}
                                \ + \ i \  F_{\A\B}^{ \ \ \ ij}
\nonumber\\
\acom{ \bar \D_{\dot\A i} }{ \bar \D_{\dot\B j} }
      &=& -2i \ g_{ij}  \ \E_{\dot\A\dot\B} \ \D_{\bar z}
          \ + \ i \ F_{\dot\A\dot\B \ ij}
\nonumber\\
\com{ \D_{m} }{\D_{\A}^{ \ j} }     &=& i \ F_{m \A}^{ \ \ \ j (r)} T_{(r)}
\nonumber\\
\com{ \D_{m} }{\bar \D_{\dot\A i} } &=& i \ F_{m \dot\A i}^{ \ \ \ (r)}T_{(r)}
\nonumber\\
 \com {\D_{m}}{\D_{n}}              &=& i \ F_{mn}^{ \ \ \ \ (r)} \ T_{(r)}
\nonumber\\
  \com{ \D_{z}}{ \D_{A} } &=& i \ F_{zA}^{ \ \ \ \ (r)} \ T_{(r)} 
\nonumber\\
  \com{ \D_{\bar z}}{ \D_{A} } &=&  i \ F_{\bar z A}^{ \ \ \ \ (r)} \ T_{(r)}.
\end{eqnarray}

Here the covariant derivatives are given as
$\D_{A} =  D_{A} + i  A_{A}$
with the index $A \sim (m, \A i, \dot\A j, z, \bar z)$.
The generators $T_{(r)}$ of the $U(1)$ gauge group satisfy
$\com {T_{(r)}}{T_{(s)}} = 0$ and
the superfields $ A_{A}  =  A_{A}^{(r)} \ T_{(r)} $
represent the $U(1)$ gauge connections.
They transform in the adjoint representation under gauge transformations:

\begin{eqnarray}
\label{connection_transformtion}
 \ A_{A}  &\rightarrow&   e^{-i \Lambda} \ A_{A} \  e^{i \Lambda}
                            -i \ e^{-i \Lambda} \ D_{A} \  e^{i \Lambda}
\end{eqnarray}

It is rather unconventional that the operator $\D_{z}$ of the gauged 
central charge does not commute with all the generators - in this sense 
it is not a `good' central charge.
\\
\\
Now we set the first constraint on the gauge connections:
Any gauge connection is independent of the central charge, i.e.

\begin{eqnarray}
\label{first_connection_constraint}
   A_{A}  &=& A_{A} \ (x^{m}, \theta^{\A}_{ \ i}, \bar\theta^{\dot\A j})
\end{eqnarray}

So we have
$\Lambda = \Lambda (x^{m}, \theta^{\A}_{ \ i}, \bar\theta^{\dot\A j})$.
and therefore the gauge connection of the central charge transforms now as
$  A_{z}  \rightarrow  e^{-i \Lambda} \ A_{z} \  e^{i \Lambda}$
under gauge transformations.
The same holds for $ A_{\bar z}$.
\\
\\
As usual the field strength is given in general as

\begin{eqnarray}
\label{general_field_strength}
 F_{AB}  &=& D_{A} A_{B} \ - \ (-)^{ab} \ D_{B} A_{A}
             \ + \ i \ T_{AB}^{ \ \ \ C } \ A_{C}
\end{eqnarray}

and the corresponding Bianchi identities are

\begin{eqnarray}
\label{graded_Bianchi_identity}
 \cyclicsum{A}{B}{C} \ (-)^{ac}
 \left \{
    D_{A} \ {F}_{BC} \ + \
    i \ {T}_{AB}^{ \ \ \ E}  \ {F}_{EC}
 \right \}
  &=&  0.
\end{eqnarray}

In our specific case with one complex central charge we find
13 field strengths and 26 Bianchi identities after elimination
of the trivial ones.
To proceed we set constraints on the field strengths
to eliminate superfluous component fields.
First of all we have the {\em natural constraints}:

\begin{eqnarray}
\label{natural_constraints}
 F_{\A\B}^{ \ \ \ ij} \ = \ F_{\A\dot\B \ \ \ j}^{ \ \ \ i} \ = \
 F_{\dot\A\dot\B \ ij}  = 0
\end{eqnarray}

These constraints are natural in the sense that
they can be obtained by an appropriate
redefinition of the gauge connections appearing in the $N=2$
algebra. Note that these constraints survive the
truncation to a $N=1$ supersymmetric theory and lead in that
case to the $N=1$ vector multiplet [\ref{Wess_Bagger}].
Second we have the {\em central charge constraints}: Some of
them follow directly from (\ref{first_connection_constraint}).

\begin{eqnarray}
\label{central_charge_constraints}
 F_{z \bar z} \ = \ F_{\dot\A \bar z \ i}  \ = \
 F_{\A z}^{ \ \ i}                                  &=& 0
\nonumber\\
 F_{az} - \partial_{a} A_{z}                        &=& 0
\nonumber\\
 F_{a \bar z} - \partial_{a} A_{\bar z}             &=& 0
\nonumber\\
 F_{\A\bar z}^{ \ \ \ i} - D_{\A}^{ \ i} A_{\bar z} &=& 0
\nonumber\\
 F_{\dot\A z \ i} - \bar D_{\dot\A \ i} A_{z}       &=& 0
\end{eqnarray}

Moreover the number of non-trivial field strengths reduces to 7
and the number of non-trivial constrained Bianchi identities
to 21. However, 10 bianchi identities are still fulfilled by
the use of the algebra. The other 11 Bianchi identities must
be solved subject to the set of constraints. The solution
yields the following constraints on the connection of the central charge:

\begin{eqnarray}
\label{non_trivial_constraint}
     D_{\A}^{ \ i} \ A_{z}  &=& 0
\\
     \bar D_{\dot\A  i} \ A_{\bar z}  &=&  0
\\
     \com{ D^{\A i} }{ D_{\A}^{ \ j} } \                   A_{\bar z}
     &=&
     \com{ \bar D_{\dot\A}^{ \ i} }{ \bar D^{\dot\A j} } \ A_{z}
\\
     \com{ D^{\A}_{ \ i} }{ D_{\A j} } \                   A_{\bar z}
     &=&
     \com{ \bar D_{\dot\A i} }{ \bar D^{\dot\A}_{ \ j} } \ A_{z}
\end{eqnarray}

All the other component fields can be
expressed in terms of the gauge connection of the central charge.
In this context the following superfield-identities hold:

\begin{eqnarray}
F_{\B \ \A\dot\A}^{ \ \ \ \ \ \ j} &=&
      \E_{\B\A} \ \bar D_{\dot\A \ i}  \ g^{ij} \ A_{z}
\\
F_{\dot\B \ \A\dot\A \ j}    &=&
      \E_{\dot\B\dot\A} \ D_{\A}^{ \ i} \ g_{ij} \ A_{\bar z}
\\
F_{\A\dot\A \ \B\dot\B}      &=&
      \E_{\A\B} \ f_{(\dot\A\dot\B)} \ + \
      \E_{\dot\A\dot\B} \ f_{(\A\B)}
\end{eqnarray}

with the definitions

\begin{eqnarray}
f_{(\A\B)}         &=&  - \ \frac{i}{4} \  D_{\A}^{ \ i} \ g_{ij} \
                        D_{\B}^{ \ j} \ A_{\bar z}
\\
f_{(\dot\A\dot\B)} &=&  - \ \frac{i}{4} \ \bar D_{\dot\A  i} \ g^{ij} \
                        \bar D_{\dot\B  j} \ A_{z}.
\end{eqnarray}

Collecting this information about the gauge connection of the
central charge we find that it is a $N=2$ vector multiplet
[\ref{Grimm_Wess_Sohnius},\ref{Firth_Jenkins}, \ref{Fayet},
\ref{deWit},\ref{Mueller}].
Note that another set of constraints might yield a different
multiplet.

\vspace{1cm}

The $8_{B} + 8_{F}$ vector multiplet
contains a complex scalar, a vector,
a left-handed $SU(2)_{R}$
spinor doublet and a real $SU(2)_{R}$ triplet, which represents
three auxiliary fields:

\begin{eqnarray}
\label{vector_multiplet_1}
A_{\bar z} & \sim & ( \ X_{\bar z}  \ , \ a_{m}
                     \ | \ \lambda_{\A \bar z}^{ \ \ \ i}
                     \ || \ Y^{ij} \ )
\end{eqnarray}

In the following we will refer to the complex scalar as the
Higgs field, whereas the spinor doublet represents the
Higgsino and the Gaugino.
The $SU(2)_{R}$ triplet obeys the constraint:

\begin{eqnarray}
\label{aux_constraint}
Y^{ij} \ = \ Y^{ji} \ = \ \bar Y_{ij}
\end{eqnarray}

The vector multiplet is defined at component
level as

\begin{eqnarray}
\label{comp_level_vector_multiplet}
 A_{\bar z |} &=& X_{\bar z} \ (x)
\\
 \frac{i}{8} \E^{\C\B} \Sm_{mn\B}^{ \ \ \ \ \A}
 D_{\A}^{ \ i} g_{ij} D_{\C}^{ \ j} A_{\bar z}
  +
 \frac{i}{8} \bar \Sm_{mn \ \ \dot\A}^{ \ \ \ \dot\B}
 \E^{\dot\A\dot\C}
 \bar D_{\dot\B \ i} g^{ij}
 \bar D_{\dot\C \ j}  A_{z |} &=& f_{mn} \ (x)
\\
 D_{\A}^{ \ i} \ A_{\bar z |} &=&  \lambda_{\A \bar z}^{ \ \ \ i} \ (x)
\\
 \com{ D^{\A i} }{ D_{\A}^{ \ j} } \   A_{\bar z |} &=& Y^{ \ ij} \ (x).
\end{eqnarray}

The field strength of the abelian component vector field
is defined in the usual way:
$f_{mn} = \partial_{m} a_{n} - \partial_{n} a_{m}$.
The component fields transform under supersymmetry transformations
generated by the operator
$\delta = \xi^{\A}_{ \ i} D_{\A}^{ \ i} +
          \bar\xi_{\dot\A i} \bar D^{\dot\A i}$ as follows:

\begin{eqnarray}
\label{susy_trafos}
 \delta X_{\bar z} &=& \xi^{\A}_{ \ i} \lambda_{\A\bar z}^{ \ \ \ i}
\\
 \delta a_{m} &=& \frac{1}{2} \ \bar\Sm_{m}^{\ \ \dot\A\A}
                  ( \xi_{\A}^{ \ i} \bar\lambda_{\dot\A z i}
                   + \bar\xi_{\dot\A i} \lambda_{\A \bar z}^{ \ \  i}
                  )
\\
\delta\lambda_{\A\bar z}^{ \ \ \ i} &=&
         \frac{1}{8} \xi^{\B}_{ \ j} \E_{\B\A} Y^{ji} -
         \frac{i}{2} \xi^{\B}_{ \ j} g^{ji} (\Sm^{mn}\E)_{\B\A} f_{mn} -
         2i \bar\xi_{\dot\A j} \E^{\dot\A\dot\B} g^{ji} \Sm^{m}_{\A\dot\B}
         \partial_{m} X_{\bar z}
\\
\delta Y^{ij} &=& 4i \xi^{\A}_{ \ k} \
 ( g^{ki} \Sm^{m}_{\A\dot\B} \partial_{m} \bar\lambda^{\dot\B \ \ j}_{ \ z} +
   g^{kj} \Sm^{m}_{\A\dot\B} \partial_{m} \bar\lambda^{\dot\B \ \ i}_{ \ z}
 )
\nonumber\\ & &
                 - 4i \bar\xi_{\dot\A k} \
  ( g^{ki} \bar\Sm^{m \dot\A\B} \partial_{m} \lambda_{\B \bar z}^{ \ \ j} +
    g^{kj} \bar\Sm^{m \dot\A\B} \partial_{m} \lambda_{\B \bar z}^{ \ \ i} )
\end{eqnarray}

The Bogomol'nyi bound of massive particle states is associated
with the central charge. Gauging the central charge can change
the bound.
To show this the inequality for the masses will be derived in a
way described in [\ref{Witten_Olive}]: First Majorana spinors
$\D^{i} = \left ( \begin{array}{c}
                    \D_{\A}^{ \ i} \\
                     \bar\D^{\dot\A i}
                   \end{array}
          \right )
$
and $\D_{j}^{ \ +} = \left (
                  \bar\D_{\dot\B i}, \D^{\B}_{ \ j}
              \right )
    $
are introduced. Then the eigenvalues of the $8 \times 8$ matrix
$\acom{\D^{i}}{\D_{j}^{ \ +}}$ must be calculated by the use
of the algebra in the rest frame. With
$i \D_{\A\dot\B} = - \delta_{\A\dot\B} M$  we
find that the eigenvalues of the matrix are real if and only if

\begin{eqnarray}
\label{bogomolnyi_bound}
 M^{2} &\geq& |\partial_{z} \ + \  i \ X_{z}|^{2}
\end{eqnarray}

Of course the eigenvalues of the matrix must be real and that is why
the inequality must hold for massive particle states.
So a vacuum expectation value of the Higgs field can change the mass bound.
\\
Changing the basis and introducing the $Q_{\A}^{ \ i}$ as supersymmetry
generators with

\begin{eqnarray}
 \acom{ Q_{\A}^{ \ i} }{ \bar Q_{\dot\B j} }
                            &=& 2 \ \Sm_{\A\dot\B}^{ \ \ m}
                                \delta^{i}_{ \ j} P_{m}
\hspace{0,5cm} \mbox{and} \hspace{0,5cm}
 \acom{ Q_{\A}^{ \ i} }{ Q_{\B}^{ \ j} }
                             \ = \ 2 \ g^{ij} \ \E_{\A\B} \ Z
\end{eqnarray}

we recover the well-known result of Olive and Witten [\ref{Witten_Olive}].
However, in this basis the effect of gauging the central charge is
hidden.


\vspace{1cm}

To conclude, it is shown in this paper that the central charge
appearing in the $N=2$ algebra in four dimensions can be gauged
by abelian gauge groups. 
\\
The case for one $U(1)$ gauge group has been studied explicitly
by solving the Bianchi identities subject to a set of constraints.
It turns out that the gauge connection of the central charge
is a $N=2$ vector multiplet with respect to these constraints.
For more than one central charge one has additional constraints
on the field strengths like
$F_{z_{i}z_{j}} = F_{\bar z_{i} \bar z_{j}} = 0$,
where the index $i$ labels the number of different central charges.
However, all the associated gauge connections of the central charges
are $N=2$ vector multiplets.  
\\
Then the effect of the gauging concerning the Bogomol'nyi bound
of the massive particle states has been investigated. It is shown that
a non-vanishing vacuum expectation value of the Higgs field can change
this mass bound. As a consequence one can understand from an algebraic
point of view, why the breaking of a non-abelian gauge group down to
$U(1)$ can change the $N=2$ central charge. Just because of the fact
that any abelian field strength $F_{\A\B}^{ \ \ \ ij}$ appearing in the
algebra

\begin{eqnarray}
\label{n=2susy_algebra3}
 \acom{ \D_{\A}^{ \ i} }{ \D_{\B}^{ \ j} }
                            &=& -2i \ g^{ij} \ \E_{\A\B} \ \D_{z}
                                \ + \ i \ F_{\A\B}^{ \ \ \ ij}
\end{eqnarray}

can enter the gauged central charge.
This lead to the {\em natural constraints}.
\\
The gauged central charge operator does not commute with all the
generators of the theory. This problem cannot be solved
in this simple, linearized appoach to supersymmetry. 
The solution of this problem might be a non-linear realization
of $N=2$ supersymmetry or the inclusion of gravity. Anyway,
problems with the central charge are well known in $N=2$ supersymmetric
gauge theories: The discussion of hypermultiplets, for example, 
leads off-shell to an infinite set of auxiliary fields. 
\\
Finally, it should be mentioned that a $N=2$ supersymmetric theory
in four dimensions can be obtained by dimensional reduction
of a $N=1$ supersymmetric theory in six dimensions
[\ref{Brink},\ref{Sohnius_Stelle_West}]. From this point of view
the generator of the complex ungauged central charge is related
to partial derivatives with respect to the two internal coordinates.
Analogous the Higgs field represents two degrees of freedom of
the six-dimensional $U(1)$ gauge connection in four dimensions.
And the constraint (\ref{first_connection_constraint}) makes
any four-dimensional gauge connection independent of the internal
coordinates.
The important difference between the six and the four dimensional
theory is that the six-dimensional $U(1)$ gauge connection cannot have
a non-vanishing vacuum expectation value. This would violate Lorentz
invariance. On the other hand in the four dimensional case the two degrees
of freedom of the six-dimensional $U(1)$ gauge connection, which
enter the Higgs field, can have a non-vanishing vacuum expectation value.


\vspace{0.3cm}

{\bf Acknowledgement:} I would like to thank S. Ketov and C. Preitschopf
for discussions about the central charge and B. de Wit
for helpful correspondence.

\vspace{0.3cm}

%
%

\section*{References}
\begin{enumerate}
\item
\label{Seiberg_Witten}
N. Seiberg and E. Witten, Nucl. Phys. {\bf B426} (1994) 19.
{\em Electric-magnetic duality, monopole condensation and confinement
in $N=2$ supersymmetric Yang-Mills theory}
\\
\\
N. Seiberg and E. Witten, Nucl. Phys. {\bf B431} (1994) 484.
{\em Monopoles, duality and chiral symmetry breaking in $N=2$
supersymmetric QCD }
\item
\label{Montonen_Olive}
C. Montonen and D. Olive, Phys. Lett. {\bf 72B} (1977) 117.
{\em Magnetic monopoles as gauge particles ?}
\item
\label{Theisen_Klemm}
A. Klemm, W. Lerche, S. Theisen and S. Yankielowicz,
 Phys. Lett. {\bf 344B} (1995) 169.
{\em Simple singularities and $N=2$ supersymmetric Yang-Mills theory}
\\
\\
A. Klemm, W. Lerche and S. Theisen,
hep-th (9505150).
{\em Non-perturbative effective actions of $N=2$ supersymmetric
gauge theories}
\item
\label{Argyres_Faraggi}
P. Argyres and A. Faraggi,
 Phys. Rev. Lett. {\bf 74} (1995) 3931.
{\em The vacuum structure and spectrum of $N=2$ supersymmetric $SU(N)$
gauge theory}
\item
\label{Luest}
B. de Wit, V. Kaplunovsky, J. Louis and D. L\"ust,
 Nucl. Phys.  {\bf B451} (1995) 53.
{\em Perturbative couplings of vector multiplets in
 $N=2$ heterotic string vacua}
\\
\\
G. Lopes Cardoso, D. L\"ust and T. Mohaupt,
Nucl. Phys.  {\bf B455} (1995) 131.
{\em Non-perturbative monodromies in $N=2$ heterotic string vacua}
\item
\label{Antoniadis}
I. Antoniadis, S. Ferrara, E. Gava, K.S. Narain and T.R. Taylor,
 Nucl. Phys.  {\bf B447} (1995) 35.
{\em Perturbative prepotential and monodromies in $N=2$ heterotic superstring}
\item
\label{Witten_Olive}
D. Olive and E. Witten, Phys. Lett {\bf 78B} (1978) 97.
{\em Supersymmetry algebras that include topological charges}
\item
\label{Haag}
R. Haag, J.T. Lopusz\'anski and M. Sohnius, Nucl. Phys.{\bf B88} (1975) 257.
{\em All possible generators of supersymmetries of the S-matrix}
\item
\label{Grimm_Wess_Sohnius}
R. Grimm, M. Sohnius and J. Wess, Nucl. Phys. {\bf B133} (1978) 275.
{\em Extended supersymmetry and gauge theories}
\item
\label{gauged_central_charge}
C.K. Zachos, Phys. Lett.  {\bf 76B}  (1978) 329.
{\em $N=2$ supergravity theory with a gauged central charge}
\\
\\
B. de Wit, J.W. van Holten and A. Van Proeyen,
Phys. Lett.  {\bf 95B}  (1980) 51.
{\em Central charges and conformal supergravity}
\\
\\
P. Claus, B. de Wit, M. Faux, B. Kleijn, R. Siebelink and P. Termonia,
hep-th 9512143.
{\em The vector-tensor supermultiplet with gauged central charge}
\item
\label{Ferrara_Zumino}
S. Ferrara and B. Zumino, Nucl. Phys. {\bf B79}  (1974) 413.
{\em Supergauge invariant Yang-Mills theories}
\item
\label{Firth_Jenkins}
R.J. Firth and J.D. Jenkins, Nucl. Phys.{\bf B85} (1975) 525.
{\em Supersymmetry with isospin}
\item
\label{Fayet}
P. Fayet, Nucl. Phys. {\bf B113}  (1976) 135.
{\em Fermi-Bose Hypersymmetry}
\item
\label{deWit}
M. de Roo, B. de Wit, J.W. van Holten and A. Van Proeyen,
 Nucl. Phys. {\bf B173} (1980) 175.
{\em Chiral superfields in $N=2$ supergravity}
\\
\\
B. de Wit, J.W. van Holten and A. Van Proeyen, Nucl. Phys. {\bf B184} (1981)
77, E {\bf B222} (1983) 516.
{\em Structure of $N=2$ supergravity}
\\
\\
E. Cremmer, J.P. Derendinger, C. Kounnas, B. de Wit,
S. Ferrara, L. Girardello and A. Van Proeyen,
Nucl. Phys. {\bf B250} (1985) 385.
{\em Vector multiplets coupled to $N=2$ supergravity:
Super-Higgs-Effect, flat potentials and geometric structure}
\\
\\
B. de Wit, P.G. Lauwers and A. Van Proeyen, Nucl. Phys. {\bf B255} (1985) 596.
{\em Lagrangians of $N=2$ supergravity-matter systems}
\item
\label{Mueller}
M. M\"uller, Nucl. Phys. {\bf B289} (1987) 557.
{\em Chiral actions for minimal $N=2$ supergravity}
\item
\label{Brink}
L. Brink, J. Schwarz and J. Scherk, Nucl. Phys. {\bf B121} (1977) 77.
{\em Supersymmetric Yang-Mills theory}
\item
\label{Sohnius_Stelle_West}
M.F. Sohnius, K.S. Stelle and P.C. West, Nucl. Phys. {\bf B173} (1980) 127.
{\em Dimensional reduction by legendr\'e transformation generates off-shell
supersymmetric Yang-Mills theory}
\item
\label{Wess_Bagger}
J. Bagger and J. Wess, Supersymmetry and Supergravity, Princeton
University Press.
\end{enumerate}

\end{document}